\documentclass{article}
\usepackage{graphicx}
\begin{document}
\title{Comment on "Indefinitely Flat Circular Velocities and the Baryonic
Tully-Fisher Relation from Weak Lensing"}
\author{Davor Palle \\
ul. Ljudevita Gaja 35, 10000 Zagreb, Croatia \\
email: davor.palle@gmail.com}
\maketitle
\begin{abstract}
{
The recent measurements of circular velocity curves from weak lensing 
of the isolated galaxies
lead to a conclusion that the circular velocity curves remain flat 
well beyond the virial radii of dark matter halos up to 1 Mpc.
This is in clear contradiction with the LCDM numerical simulations.
We show that the additional cosmic force originating from the geometry
of the Universe beyond the Friedmann model increases the circular 
velocities of the test galactic bodies and avoids a decline of the
standard LCDM velocity curves.
}
\end{abstract}

We noticed very interesting measurements of the circular velocity
curves of the early and late time isolated galaxies with very
disturbing conclusion contradicting the standard LCDM galactic halo
numerical simulations \cite{Mistele}. Instead of the decline of the velocity 
curves at large distances, they established a flat circular velocity curves up to 
1 Mpc from the galactic centres. It seems that this characteristic is 
universal with a significant statistical evidence.

Since the LCDM cosmology proved to be a successful theoretical concept
to resolve the CMB, lensing, BAO and galactic dynamics, it is very difficult
to find a viable alternative framework. However, some recent anomalies of the 
LCDM like Hubble tension, $S_{8}$ tension, EDGES measurements or the
very surprising high redshift Universe revealed by JWST are directing
our research toward the Einstein-Cartan cosmology \cite{Palle1}.
The inevitable appearance of the vorticity and acceleration of the primordial
origin turns out to be closely related to torsion of spacetime and 
the particle content of the Universe defined by the nonsingular particle
theory with the heavy Majorana neutrinos as the CDM particles accompanied
with the light Majorana neutrinos as the HDM particles \cite{Palle2}.

We demonstrated the importance of vorticity in our paper on the pulsar
timing arrays \cite{Palle3} with no need to 
reference to gravity waves. It is well known that the Hubble expansion
effect is present even in the analysis of the Lunar Laser Ranging
experiment \cite{Maeder}. Analogously, the cosmic acceleration can have
its imprint in the solar system \cite{Palle4} as an additional cosmic
force in the direction toward the Sun. We can use the same imbedding
formalism of Gautreau to claim for the additional
acceleration of the cosmological origin toward the galactic centres.

The metric used in the Einstein-Cartan cosmology \cite{Palle1,Palle3}
receives the following form:

\begin{eqnarray}
d s^{2}=d t^{2}-R^{2}(t)[d x^{2}+(1-\lambda^{2}(t))e^{2mx}d y^{2}
+d z^{2}]-2 R(t)\lambda(t) e^{mx}d y dt, \\
m=const., \nonumber
\end{eqnarray}

The Ehlers decomposition is:

\begin{eqnarray}
\nabla_{\mu}u_{\nu}=\omega_{\nu\mu}
+\sigma_{\mu\nu}+\frac{1}{3}\Theta h_{\mu\nu}+u_{\mu}a_{\nu}, \\
u^{\mu}u_{\mu}=1,\ h_{\mu\nu}=g_{\mu\nu}-u_{\mu}u_{\nu},
\ a_{\mu}=u^{\nu}\nabla_{\nu}u_{\mu}, 
\ \Theta=\nabla_{\nu}u^{\nu}, \nonumber \\
\omega_{\mu\nu}=h^{\alpha}_{\mu}h^{\beta}_{\nu}
\nabla_{[\beta}u_{\alpha ]},\ 
\sigma_{\mu\nu}=h^{\alpha}_{\mu}h^{\beta}_{\nu}
\nabla_{(\alpha}u_{\beta )}-\frac{1}{3}\Theta h_{\mu\nu},
\nonumber 
\end{eqnarray}
\begin{eqnarray*}
  [\alpha\beta]\equiv\frac{1}{2}(\alpha\beta-\beta\alpha),\ 
(\alpha\beta)\equiv\frac{1}{2}(\alpha\beta+\beta\alpha). \nonumber
\end{eqnarray*}

In this work we neglect the vorticity ($m=0$, c=velocity of light):

\begin{eqnarray}
H\equiv\frac{\dot{R}}{R},\ \omega\equiv(\frac{1}{2}\omega_{\mu\nu}\omega^{\mu\nu})^{1/2}=
\frac{m \lambda}{2 R}c=0,\ a\equiv(-a_{\mu}a^{\mu})^{1/2}=(\dot{\lambda}+\lambda\frac{\dot{R}}{R})c.
\end{eqnarray}

We describe galactic halo by the Navarro, Frenk and White (NFW) \cite{Navarro}
spherical density profile:

\begin{eqnarray}
\rho(r) = \frac{\rho_{s} r^{3}_{s}}{r (r+r_{s})^{2}}.
\end{eqnarray}

It is parametrized from the numerical simulations within the CDM framework.
The resulting circular velocity is \cite{Navarro}:

\begin{eqnarray}
v^{2}(r) = \frac{G_{N} M_{*}}{r}[\ln(1+x)-\frac{x}{1+x}], \\
x=r/r_{s},\ M_{*}=4 \pi \rho_{s} r_{s}^{3}.
\end{eqnarray}

The maximum velocity $v_{max}$ is achieved at $x \simeq 2.16$.

With the addition of the universal cosmic acceleration, the galactic
velocity profile acquires the form:

\begin{eqnarray}
v_{tot}^{2}(r)=\frac{G_{N} M_{*}}{r}[\ln(1+x)-\frac{x}{1+x}]
+a r.
\end{eqnarray}

We use the asymptotic form of the weak gravity limit at large
distances:

\begin{eqnarray}
v_{tot}^{2}(x)=v_{halo}^{2}\frac{\ln x}{x} + v_{a}^{2} x .
\end{eqnarray}

Imposing the condition of the nondeclining velocities within the distances
100 kpc-1000 kpc from the observation of Mistele et al. \cite{Mistele}:

\begin{eqnarray}
v_{tot}(x_{1}) = v_{tot}(10 x_{1}),\ x_{1}=r_{1}/r_{s},\ r_{1}=100\ kpc,
\end{eqnarray}

we find for the ratio:

\begin{eqnarray}
\frac{v_{a}^{2}}{v_{halo}^{2}}=\frac{10\ln x_{1}-\ln (10 x_{1})}
{90 x_{1}^{2}}.
\end{eqnarray}

For a velocity $v_{tot}(x_{1})=250 km/s$ we can evaluate $v_{halo}$:

\begin{eqnarray}
v_{halo}^{2}=v_{tot}(x_{1})^{2}[\ln x_{1}/x_{1}+\frac{v_{a}^{2}}{v_{halo}^{2}}
x_{1}]^{-1}
\end{eqnarray}

Since a typical $r_{max}\simeq 10 kpc$, one can finally estimate 
the universal cosmic acceleration (obviously it depends generally on the redshift
a=a(z)):

\begin{eqnarray}
r_{s}=r_{max}/2.16,\ a = \frac{v_{a}^{2}}{r_{s}}=1.70\times 10^{-12}m s^{-2}.
\end{eqnarray}

The reader can envisage asymptotic velocity profiles in Fig.1.

\begin{figure}[htb]
\centerline{
\includegraphics[width=12cm]{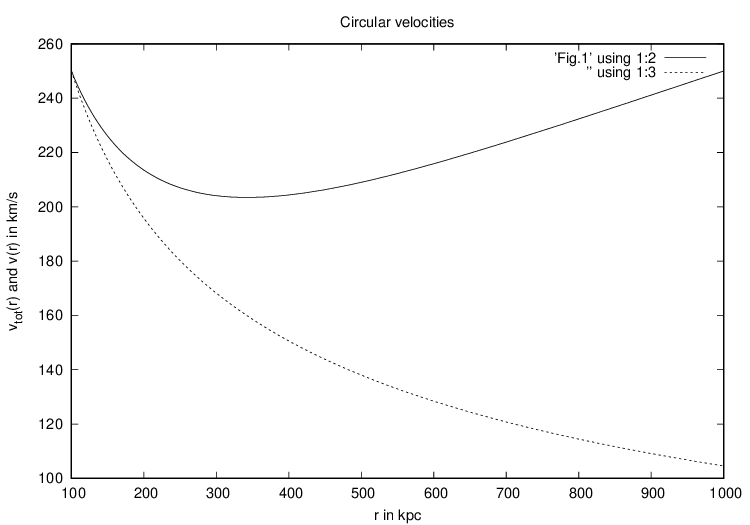}}
\caption{Solid curve is $v_{tot}$(r) and a dashed one v(r).} 
\end{figure}

We can conclude that the additional cosmological acceleration 
improves the behaviour of the CDM velocity curves at very large
distances in accordance with the observations.
The cosmic acceleration could play also the important role in dynamics of
wide binary stars \cite{Xavier}.

\end{document}